\documentstyle[aps,prd,multicol]{revtex}
\tighten

\begin{document}

\title{Hamiltonian formulation of nonAbelian noncommutative gauge theories}

\author{Ricardo Amorim$^a$ and Franz A. Farias$^b$}

\address{\mbox{}\\
Instituto de F\'{\i}sica\\
Universidade Federal do Rio de Janeiro\\
RJ 21945-970 - Caixa Postal 68528 - Brasil}
\date{\today}

\maketitle
\begin{abstract}
\hfill{\small\bf Abstract\hspace*{1.7em}}\hfill\smallskip
\par
\noindent
We implement the Hamiltonian treatment of a nonAbelian noncommutative gauge theory, considering with some detail the algebraic structure of the noncommutative symmetry group. The first class constraints and Hamiltonian are obtained and
their algebra derived, as well as the form of the gauge invariance they impose on the first order action.
\end{abstract}

\smallskip\mbox{}

\section{Introduction}
\renewcommand{\theequation}{1.\arabic{equation}}
\setcounter{equation}{0}

In recent years  there has been a great interest in noncommutative field theories. This is due not only because of their features, which constitute remarkable generalizations of those presented by conventional field theories, but also because they naturally appear in the context of string theories\cite{SW}.  
To construct the noncommutative version of a field theory one basically
replace the product of fields in the action by the Moyal product:

\begin{equation}
\phi_1(x)\star\phi_2(x)=\exp\,
\left(\frac{i}{2}\theta^{\mu\nu}\partial_\mu^x\partial_\nu^y\right)\,
\phi_1(x)\phi_2(y)\vert_{x=y}
\label{1.1}
\end{equation}

\noindent
where $\theta^{\mu\nu}$ is a real and antisymmetric constant matrix.
As can be verified, the space-time integral of the Moyal product of two fields is the same as the  usual one, provided we discard boundary
terms. So the noncommutativity affects just the vertices in the action ( see  Appendix A, where some properties  of the Moyal product are listed ).
\medskip

Regardless the enormous amount of works written on the subject, only a few
of them concern the Hamiltonian treatment of noncommutative theories. They
consider both the cases depending if the noncommutative parameter $\theta^{0i}$ vanishes \cite{Dayi,AB1} or not \cite{G1,AB2}. In the second situation we  necessarily face an arbitrarily higher order derivative theory and it has to be treated with non canonical means. The examples found in literature consider only the Hamiltonian formulation of  noncommutative $U(1)$ gauge theories. In a more general setting, however, formally the Lie commutators 
of the corresponding nonAbelian commutative theory are replaced by Moyal commutators. This modification implies, among other features, that $SU(N)$ can not consistently be the symmetry  group of a noncommutative action\cite{SW}. This has remarkable consequences, not only related with the structure of the correspondence between commutative and noncommutative gauge theories \cite{SW,Wess}, but also
implies severe restrictions over the phenomenology described by the theory \cite{B,J1}.
 \medskip

In the present work we will consider the Hamiltonian treatment of a general nonAbelian noncommutative gauge theory adopting the condition $\theta^{0i}=0$, to keep unitarity \cite{G2} and avoid non canonical means. In Section II we give  a brief review of the subject, necessary to stablish conventions and notations,  and after that we discuss the enveloping algebra structure of the theory and its relation with the invariance under the $U(N)$ symmetry group. It is then possible to implement the Hamiltonian treatment of the noncommutative gauge theory, displaying constraints, Hamiltonian,
their first class algebra and the gauge invariance of the corresponding first order action, which is done in Section III. All of this strongly depends on the fact that connections, curvature, parameters etc take values in an algebra that closes not only under commutation but also under anticommutation. Although we are here adopting a commutative time, the algebraic structure introduced by the Moyal product is not trivial. For instance, we only can prove the closure of the constraints algebra by using evolved expressions that come from the noncommutative version of Jacobi identity, as can be verified from the developments of Section III. We reserve Section IV for some concluding remarks.\bigskip

\section{$u(N)$ versus enveloping gauge algebra }
\renewcommand{\theequation}{2.\arabic{equation}}
\setcounter{equation}{0}

It is well known that the connections of noncommutative gauge theories can not consistently take values in any $su(N)$ algebra, but  in a $u(N)$ one. As already commented, there are also  formulations of noncommutative gauge theories that extensively employ the concept of enveloping algebras, whose generators in principle would be formed by all the  products, in all orders, of some original set of generators. We discuss in this section a connection between both descriptions, which will be show itself to be essential for the developments presented in the rest of the present work.
Let us start by considering 
the action which describes the gauge sector of a noncommutative nonAbelian theory, which can be written as 

\begin{equation}
S=-\,\frac{1}{2}\, tr \int d^4x\,F_{\mu\nu}\star F^{\mu\nu}
\label{2.1}
\end{equation}

\noindent where the curvature tensor is defined by

\begin{eqnarray}
F_{\mu\nu}&=&\partial_\mu A_\nu-\partial_\nu A_\mu
-i\,(A_\mu\star A_\nu-A_\nu\star A_\mu)
\nonumber\\
&\equiv&\partial_\mu A_\nu-\partial_\nu A_\mu
-i\,[A_\mu,A_\nu]
\label{2.2}
\end{eqnarray}

 Action (\ref{2.1}) is invariant under the gauge transformations

\begin{eqnarray}
\delta A_\mu&=&D_\mu\Lambda\nonumber\\
&=&\partial_\mu\Lambda-i[A_\mu,\Lambda]
\label{2.3}
\end{eqnarray}

\noindent
since under (\ref{2.3})

\begin{equation}
\delta F_{\mu\nu}=-i[F_{\mu\nu},\Lambda]
\label{2.4}
\end{equation}

\noindent and the Moyal product is associative and  satisfies the cyclic property under the integral sign.
It is also easy to deduce that transformations (\ref{2.3}) close in an algebra

\begin{equation}
[\delta_1,\delta_2]A_\mu=\delta_3 A_\mu
\label{2.5}
\end{equation}

\noindent where

\begin{equation}
\Lambda_3=i[\Lambda_1,\Lambda_2]
\label{2.6}
\end{equation}

 The above expressions are deduced by using the Jacobi identity  (A6) and the associativity of the Moyal product, without assuming any further
details of the algebraic structure constraining the gauge connections. However, the gauge transformations (\ref{2.3}) imply that the gauge fields cannot take values, for instance, in a $su(N)$ algebra, but  in some section of the corresponding enveloping algebra \cite{Wess}.
This is an essential feature of noncommutative nonAbelian gauge theories, with several fundamental implications. To understand the origin of this fact, suppose that in a first approximation, $A_\mu =A_\mu^a\,T^a$ and $\Lambda =\Lambda^a\,T^a$, where
the  $T$'s are the hermitian generators of some Lie algebra $g$ 
in some representation $R$, and  satisfying the relations

\begin{equation}
[T^a,T^b]=if^{abc}T^c
\label{2.7}
\end{equation}

As the $T$'s are constant, it is obvious that in the above expressions it is employed the usual ( Lie) commutator. Now, from  ({\ref{2.3}), it follows that

\begin{eqnarray}
\delta A_\mu&=&\partial_\mu\Lambda^a\,T^a-iA^a_\mu*\Lambda^b \,T^aT^b+i\Lambda^b*A^a_\mu\, T^bT^a
\nonumber\\
&=&\partial_\mu\Lambda^a\,T^a-{i\over2}[A^a_\mu,\Lambda^b]\,\{ T^a,T^b\}-
{i\over2}\{A^a_\mu,\Lambda^b\}\,[ T^a,T^b]
\label{2.8}
\end{eqnarray}

\noindent and so $A_\mu$ is forced to take values not only along $T^a$ but also along 
$\{T^a,T^b\}$. Iterating this procedure, we can easily be convinced that the gauge fields can take values
along the structure formed by the products, in all orders, of the Lie generators $T^a$.
In constructing the independent generators, we are free to use the  Lie algebra structure given by (\ref{2.7}). When the $T^a$'s are the generators of some Lie algebra $g$ in some representation $R$, we call this section of the corresponding enveloping algebra as $u(g,R)$, where the (anti)commutators are constructed with  usual matrix products.  In our formulation, connections, curvatures, gauge parameters etc take values in $u(g,R)$. It is easy to verify that the generators of $u(g,R)$ can be written as

\begin{equation}
T^A=(T^a,{1\over2}\{T^a,T^b\},{1\over4}\{T^a,\{T^b,T^c\}\},...)
\label{2.9}
\end{equation}

\noindent where the range of the index $A$ depends on $u(g,R)$. For example, suppose that $u(g,R)$ is given by $su(2)$ in a representation  $R$ constructed with the Pauli matrices $\sigma^i$. Then it follows that $u(g,R)$ is four dimensional and  spanned by the Pauli matrices themselves and by the two by two unit matrix, since the terms obtained by higher order anticommutators are linearly dependent of those. Obviously this is not a faithful representation of the infinite dimensional universal enveloping algebra of $su(2)$ but gives the standard representation of $u(2)$. In the adjoint representation of $su(2)$, where $(T^i)^{jk}=-i\epsilon^{ijk}$, it is easy to see that the process described above gives 
nine $3\times3$ linearly independent hermitian matrices, and so we get the standard  representation of $u(3)$.   \medskip

In general, we verify that the generators of $u(g,R)$ given by (\ref{2.9}) not only form a Lie algebra  but  also close under anticommutation:

\begin{eqnarray}
[T^A,T^B]&=&if^{ABC}T^C\nonumber\\
\{T^A,T^B\}&=&d^{ABC}T^C
\label{2.10}
\end{eqnarray}

\noindent where $f^{ABC}=-f^{BAC}$ and $d^{ABC}=d^{BAC}$. The simpler nontrivial algebra that matches these conditions is $u(N)$ in the representation given by $N\times N$ hermitian matrices. Following \cite{B}, one can choose $T^0={1\over\sqrt{2N}}\mathbf1_{NxN}$  and the remaining $N^2-1$ of the $T$'s as in $su(N)$. It is then possible to use the trace condition

\begin{equation}
tr(T^AT^B)={1\over2}\delta^{AB}
\label{2.11a}
\end{equation}

\noindent and take $f^{ABC}$ and $d^{ABC}$ as  completely antisymmetric and completely symmetric respectively. From now on we will assume these conditions.
\bigskip

Now
we can explicitly write the gauge connection and the curvature as 

\begin{eqnarray}
A_\mu&=&A_\mu^A T^A\nonumber\\
F_{\mu\nu}&=&F_{\mu\nu}^AT^A
\label{2.11}
\end{eqnarray}

\noindent and in terms of components, (\ref{2.2}),(\ref{2.10}) and (\ref{2.11})
permit us to write, for instance, that

\begin{equation}
F_{\mu\nu}=\left(\partial_\mu A^D_\nu-\partial_\nu A^D_\mu
+{1\over2}f^{BCD}\{A^B_\mu, A^C_\nu\}-{i\over2}d^{BCD}[A^B_\mu, A^C_\nu]\right)T^D
\label{2.12}
\end{equation}

\noindent Observe that in the commutative limit, the curvature components do not depend on the $d$'s, as expected. However, in the general situation, the appearance of those structure functions
is essential for the achievement of the gauge invariance of the noncommutative gauge theory. In a similar way, when we expand the gauge parameter along  the generators of $u(g,R)$, we observe that (\ref{2.3}) and (\ref{2.4}) can be rewritten as

\begin{equation}
\delta A_\mu=\left(\partial_\mu\Lambda^D+{1\over2}f^{BCD}\{A^B_\mu, \Lambda^C\}-{i\over2}d^{BCD}[A^B_\mu, \Lambda^C]\right)T^D
\label{2.13}
\end{equation}

\noindent and

\begin{equation}
\delta F_{\mu\nu}=\left({1\over2}f^{BCD}\{F_{\mu\nu}^B, \Lambda^C\}-{i\over2}d^{BCD}[F_{\mu\nu}^B, \Lambda^C]\right)T^D
\label{2.14}
\end{equation}

The notation used above, where the generators of $u(g,R)$ appear explicitly, is more evolving than that  one used, for instance, in (\ref{2.3}) and (\ref{2.4}).
However, all the calculations can be done inside this notation and it shows itself to be essential
for  implementing the Hamiltonian formulation, which will be done in what follows.
\bigskip

\section{Hamiltonian description}
\renewcommand{\theequation}{3.\arabic{equation}}
\setcounter{equation}{0}

From  (\ref{2.11a}) and (\ref{2.12}) we can write (\ref{2.1}) as

\begin{equation}
S=-\,\frac{1}{4}\,  \int d^4x\,F_{\mu\nu}^A F^{\mu\nu A}
\label{3.1}
\end{equation}

If we had adopted no restriction on the Moyal product, the Lagrangian
density appearing in (\ref{3.1}) would present time derivatives of arbitrary higher order. This would not only break unitarity\cite{G2} but also demand non trivial means to the theory be treated inside a Hamiltonian formalism. As we are considering the noncommutative parameter to satisfy $\theta_{\mu\nu}\theta^{\mu\nu}>0$ and assuming a Lorentz referential where $\theta^{0i}$ vanishes identically, we can treat action (\ref{3.1}) in a canonical way.
So we derive the momenta conjugate to $A^B_\mu$ as

\begin{eqnarray}
\Pi^B_\mu&=&{{\partial L}\over{\partial \dot A^{\mu B}}}\nonumber\\
&=&F^B_{\mu 0}
\label{3.2}
\end{eqnarray}

This means that there are primary constraints

\begin{equation}
T_1^A=\Pi^A_0
\label{3.3}
\end{equation}

\noindent and making use of a partial integration as well as of the symmetry properties of the structure functions $f^{ABC}$ and $d^{ABC}$, we arrive at the primary Hamiltonian

\begin{equation}
H_p=\int d^3x\,( {1\over2}\Pi^{iB} \Pi^{iB} +{1\over4}F_{ij}^B\, F^{ijB}-(D_i\Pi^i)^B A^{0B}+
\lambda^{1B}T_1^B)
\label{3.4}
\end{equation}

\noindent where

\begin{equation} 
(D_i \Pi^i)^B=\partial_i\Pi^{iB}+{1\over2}f^{BCD}\{A_i^C,\Pi^{iD}\}-{i\over2}
d^{BCD}[A_i^C,\Pi^{iD}]
\label{3.4b}
\end{equation}
\medskip

By using the Poisson brackets definition

\begin{equation}
\{X(x),Y(y)\}_{PB}=\int d^3z\,\biggl(
\frac{\delta X(x)}{\delta A_\mu^C(z)}
\frac{\delta Y(y)}{\delta \Pi^{\mu C}(z)}
-\frac{\delta Y(y)}{\delta A_\mu^C(z)}
\frac{\delta X(x)}{\delta \pi^{\mu C}(z)}
\biggr)
\label{3.5}
\end{equation}

\noindent where $x^0=y^0=z^0$, it is easy to see that there are secondary constraints

\begin{eqnarray}
\{T_1^A,H_p\}_{PB}&=&(D_i\Pi^i)^A\nonumber\\
&\equiv&T_2^A
\label{3.6}
\end{eqnarray}\noindent where $(D_i\Pi^i)^A$ is given by (\ref{3.4b}).
Now one can easily verify that  the constraint $T_1^A$ satisfies the abelian algebra

\begin{eqnarray}
\{T_1^A(x),T_1^B(y)\}_{PB}&=&0\nonumber\\
\{T_1^A(x),T_2^B(y)\}_{PB}&=&0\nonumber\\
\label{3.7}
\end{eqnarray}

\noindent but it is not so trivial to show that $T_2^A$ closes in an algebra with itself. Let us consider this point with some detail: From definition (\ref{3.6}), $\{T_2^A(x),T_2^B(y)\}_{PB}$ generates nine terms. Obviously
$\{\partial_i\Pi^{iA}(x),\partial_j\Pi^{jB}(y)\}_{PB}$ vanishes identically. It is also not hard to show that

\begin{eqnarray}
\{\partial_i\Pi^{iA}(x),{1\over2}f^{BCD}\{A_j^C(y),\Pi^{Dj}(y)\}\}_{PB}&+&    \{{1\over2}f^{ACD}\{A_i^C(x),\Pi^{Di}(x)\},\partial_j\Pi^{jB}(y)\}_{PB}=\nonumber\\
{1\over2}f^{ABC}(\partial_i^x\{\delta(x-y),\Pi^{iC}(y)\}&+&\partial^y_i\{\delta(x-y),\Pi^{iC}(x)\})\label{3.7b}
\end{eqnarray}

By using  (A4,5) one can see that the above expression reduces to
${1\over2}f^{ABC}\{\delta(x-y),\partial_i\Pi^{iC}(x)\}$. In a similar way, we see that

\begin{equation}
\{\partial_i\Pi^{iA}(x),{i\over2}d^{BCD}[A_j^C(y),\Pi^{Dj}(y)]\}_{PB}+    \{{i\over2}d^{ACD}[A_i^C(x),\Pi^{Di}(x)],\partial_j\Pi^{jB}(y)\}_{PB}=
-{i\over2}d^{ABC}\{\delta(x-y),\partial_i\Pi^{iC}(x)\}\label{3.7c}
\end{equation}

The terms comming from $\{T_2^A(x),T_2^B(y)\}_{PB}$ and not involving other derivatives than those comming from the Moyal (anti)commutators can be shown to sum, after some algebra, as

\begin{eqnarray}
{1\over4}f^{AXD}f^{BCX}\{\Pi_i^D,\{A^{iC},\delta\}\}&-&
{1\over4}f^{ACX}f^{BXD}\{A_i^C,\{\Pi^{iD},\delta\}\}\nonumber\\
-
{1\over4}d^{AXD}d^{BCX}[\Pi_i^D,[A^{iC},\delta]]&+&
{1\over4}d^{ACX}d^{BXD}[A_i^C,[\Pi^{iD},\delta]]\nonumber\\
+
{i\over4}f^{AXD}d^{BCX}\{\Pi_i^D,[A^{iC},\delta]\}&+&
{i\over4}f^{ACX}d^{BXD}\{A_i^C,[\Pi^{iD},\delta]\}\nonumber\\
+
{i\over4}d^{AXD}f^{BCX}[\Pi_i^D,\{A^{iC},\delta\}]&+&
{i\over4}d^{ACX}f^{BXD}[A_i^C,\{\Pi^{iD},\delta\}]
\label{3.7d}
\end{eqnarray}

\noindent where $A^{iC}=A^{iC}(x)$, $\Pi=^D_i\Pi^D_i(x)$ and $\delta=\delta(x-y)$ and all the commutators and anticommutators are calculated with respect to the Moyal product. The terms above can be written in a more convenient form. To achieve this, we observe that
if we expand  the Jacobi identity 

\begin{equation}
[\Pi_i,[A^i,\, T^B\delta]]+[A^i,[T^B\delta ,\Pi_i]]+[T^B\delta,[\Pi_i,A^i]]=0
\label{JI}
\end{equation}

\noindent by explictly writing its structure functions, 
we recognize that the eight terms in (\ref{3.7d})   simplify, with the use of (\ref{JI}), to

\begin{equation}
-{i\over2}f^{ABC}\{\delta(x-y),[A_i,\Pi^i]^C\}-{1\over2}d^{ABC}[\delta(x-y),[A^i,\Pi^i]^C\}
\label{3.7e}
\end{equation}

Collecting the expressions that come from (\ref{3.7b}), (\ref{3.7c}) and (\ref{3.7e}), we arrive at
the remaining term of the constraint algebra:

\begin{equation}
\{T_2^A(x),T_2^B(y)\}_{PB}={1\over2}f^{ABC}\{\delta(x-y),T^C_2(x)\}-
{i\over2}d^{ABC}[\delta(x-y),T^C_2(x)]
\label{3.7f}
\end{equation}

 As can be observed, the above expressions  present the correct symmetry properties
under the change $(xA)\leftrightarrow (yB)$. In a similar way, one can also prove that

\begin{eqnarray}
\{T_2^A,H\}_{PB}&=&{1\over2}f^{ABC}\{\lambda^{2B}-A^{0B},T_2^C\}-{i\over2}d^{ABC}[\lambda^{2B}-A^{0B},T_2^C]
\nonumber\\
&=&-i[\lambda^2-A^{0},T_2]^A
\label{3.8}
\end{eqnarray}

\noindent by using the appropriate Bianchi identity, and so  no more constraints are produced. In the above equation, 

\begin{equation}
\label{3.8b}
H=H_p+\int d^3x\lambda^{2A}T_2^A
\end{equation}

\noindent is the first class Hamiltonian.
\bigskip

Before concluding, let us consider the gauge invariance of the first order action

\begin{eqnarray}
S_{FO}&=&\int d^4x  \,\Pi^{\mu B}\,\dot A_\mu^B - \int dx^0 H\nonumber\\
&=&tr\int d^4x \, (2\Pi^{\mu}\,\dot A_\mu-\Pi^i\Pi^i-{1\over2}F_{ij}F^{ij}-2T_2(\lambda^2-A^{0})-2T_1\,\lambda^1) \label{3.9}
\end{eqnarray}

The gauge generator

\begin{equation}
G=\int d^3x \,(\epsilon^{1A}T_1^A+\epsilon^{2A}T_2^A)
\label{3.10}\end{equation}

\noindent acts  canonically on the phase space variables $y$ through $\delta y=\{y,G\}_{PB}$ to produce the gauge transformations

\begin{eqnarray}
\delta A^{0B}&=&\epsilon^{1B}  \nonumber\\
\delta A_i^B &=& -(D_i\epsilon^{2})^B \nonumber\\
\delta \Pi^{0B}&=&0  \nonumber\\
\delta \Pi_i^B&=&-i[\epsilon^{2},\Pi]^B  
\label{3.11}
\end{eqnarray}

It is now a simple algebraic task to show that indeed (\ref{3.9}) is invariant under (\ref{3.11}) once

\begin{eqnarray}
\delta\lambda^1&=& \dot\epsilon^1\nonumber\\
\delta\lambda^2&=&\epsilon^1+\dot\epsilon^2-i[\lambda^2-A^{0},\epsilon^2]
\label{3.12}
\end{eqnarray}

\noindent As one can observe, all of the above expressions have the proper  nonAbelian commutative  as well as the $U(1)$ noncommutative limits. It is worthwhile to notice that the use of both relations appearing in (\ref{2.10}) has been essential for deriving fundamental relations such as (\ref{3.6}), (\ref{3.7}), (\ref{3.7f}) and (\ref{3.8}). Of course, this is not the case for the commutative nonAbelian gauge theories, where the Hamiltonian treatment can be constructed only with the usual Lie algebra structure functions.
\bigskip

\section{Conclusion}

We conclude this work by remarking that we have derived a consistent Hamiltonian formulation for the gauge sector of a nonAbelian noncommutative gauge theory,
succeeding in displaying the first class constraints and Hamiltonian, their non trivial algebra  and the way they generate the evolution and gauge invariance of
phase space quantities. In doing so, it has been necessary to use several properties satisfied by field variables and distributions
when operated under the Moyal product.
Of course it remains to consider not only the inclusion of matter fields but several fundamental points related with the quantization procedure, as the extension of the phase space by the appropriate ghost fields in order to construct a $BRST$ invariant action and its corresponding path integral, with consistent gauge fixing and measure . Also the Hamiltonian extension of the Seiberg-Witten map can be studied in this formulation. These and other topics will be reported elsewhere \cite{ABF}.

\vspace{1cm}
\noindent
{\bf Acknowledgment:} We thank J. Barcelos-Neto, M. V. Cougo-Pinto and F. J. Vanhecke for inspiring discussions. This work is supported in part by  CAPES and CNPq (Brazilian research agencies).
\section*{Appendix}
\section*{Some identities related to the Moyal product}
\renewcommand{\theequation}{A.\arabic{equation}}
\setcounter{equation}{0}

\begin{equation}
\int d^4x\,\phi_1\star\phi_2=\int d^4x\,\phi_1\phi_2
=\int d^4x\,\phi_2\star\phi_1
\label{A.1}
\end{equation}

\begin{equation}
\bigl(\phi_1\star\phi_2\bigr)\star\phi_3
=\phi_1\star\bigl(\phi_2\star\phi_3\bigr)
=\phi_1\star\phi_2\star\phi_3
\label{A.2}
\end{equation}

\begin{equation}
\int d^4x \,\phi_1\star\phi_2\star\phi_3
=\int d^4x\,\phi_2\star\phi_3\star\phi_1
=\int d^4x\,\phi_3\star\phi_1\star\phi_2
\label{A.3}
\end{equation}

\begin{equation}
\phi(x)\star\delta(x-y)=\delta(x-y)\star\phi(y)
\label{A.4}
\end{equation}

\begin{equation}
\phi(x)\star\partial^x_\lambda\delta(x-y)=-\partial^y_\lambda\delta(x-y)\star\phi(y)
\label{A.5}
\end{equation}

\begin{equation}
[\phi_1,[\phi_2,\phi_3]]+[\phi_2,[\phi_3,\phi_1]]
+[\phi_3,[\phi_1,\phi_2]]=0
\label{A.6}
\end{equation}

\begin{equation}
[\phi_1(x),[\phi_2(x),\delta(x-y)]]
=[\phi_2(y),[\phi_1(y),\delta(x-y)]]
\label{A.7}
\end{equation}

\end{document}